\documentclass[superscriptaddress,showpacs,aps,prb,twocolumn]{revtex4-1}
\usepackage{times}
\usepackage{amsmath}
\usepackage{amssymb}
\usepackage{graphicx}
\usepackage{pst-all}
\usepackage{epstopdf}
\begin{document}
 
\title{Phase transition and intrinsic metric of the dipolar fermions in quantum Hall regime} 
\author{Zi-Xiang Hu}
\email{zxhu@cqu.edu.cn}
\author{Qi Li}
\author{Lin-Peng Yang}
\author{Wu-Qing Yang}
\affiliation{Department of Physics, Chongqing University, Chongqing 401331, People's Republic of China}
\author{Na Jiang}
\affiliation{Zhejiang Institute of Modern Physics, Zhejiang University, Hangzhou 310027, People's Republic of China}
\author{Rui-Zhi Qiu}
\affiliation{Science and Technology on Surface Physics and Chemistry Laboratory, Mianyang 621907, People's Republic of China}
\author{Bo Yang}
\affiliation{Complex Systems Group, Institute of High Performance Computing, A*STAR 138632, Singapore}
\date{\today}

\begin{abstract}
  For the fast rotating quasi-two-dimensional dipolar fermions in the quantum Hall regime, the interaction between two  dipoles breaks the rotational symmetry when the dipole moment has in-plane components that can be tuned by an external field.  Assuming that all the dipoles are polarized in the same direction, we perform the numerical diagonalization for finite size systems on a torus. We find that while $\nu = 1/3$ Laughlin state is stable in the lowest Landau level (LLL), it is not stable in the first Landau level (1LL); instead, the most stable Laughlin state in the 1LL is the  $\nu = 2 + 1/5$ Laughlin state. These FQH states are robust against moderate introduction of anisotropy, but large anisotropy induces a transition into a compressible phase in which all the particles are attracted and form a bound state. We show that such phase transitions can be detected by the intrinsic geometrical properties of the ground states alone. The anisotropy and the phase transition are systematically studied with the generalized pseudopotentials and characterized by the intrinsic metric, the wave function overlap and the nematic order parameter. We also propose simple model Hamiltonians for this physical system in the LLL and 1LL respectively.
\end{abstract}
 \maketitle

\section{Introduction}
Topological phase of matter and the phase transition have been the focus of much recent theoretical interests. The fractional quantum Hall (FQH) states, which are realized in two dimensional electron gas placed in strong magnetic fields, are prime examples of the strongly correlated topological system. A wide variety of the existed exotic Abelian and non-Abelian FQH states, such as the Laughlin state at filling fraction $\nu = 1/3$~\cite{Tsui-PhysRevLett.48.1559,Laughlin-PhysRevLett.50.1395} and the Moore-Read-like state at $\nu = 5/2$~\cite{PhysRevLett.59.1776, Moore1991362}, are found to host non-trivial topological properties. Their theoretical interpretations enrich our understanding of the state of the matter in condensed matter physics. The crucial point to explain the FQH effects is considering the electron-electron interaction within a magnetic field. For an ideal system with two particles, the electron-electron Coulomb interaction has the translational and rotational symmetry. However, in real materials, the rotational symmetry can be easily broken by anisotropic effective band mass~\cite{BoYangPhysRevB.85.165318,  XinWanPhysRevB.86.035122, Apalkov2014128, PhysRevLett.110.206801}, the anisotropic dielectric constant, the external strain, or the tilted magnetic field\cite{PhysRevLett.82.394, PhysRevLett.84.4685, PapicTilt, botitleB}.
Haldane ~\cite{HaldaneGeometry} pointed out that the rotational symmetry is not necessary for the FQH physics; the FQH states possess ``geometrical" degrees of freedom that are fundamental to their low-energy properties~\cite{CanLaskinWiegmann, BradlynRead, Gromov}.  For two-body interactions,  the ``geometrical" degrees of freedom can be defined by a metric characterizing the ``area preserving" quantum fluctuations of topological composite particles within a single Landau level. The notion of geometry has also inspired the construction of a more general class of FQH states with non-Euclidean metric~\cite{QiuPhysRevB.85.115308}, which were used to characterize intrinsic non-trivial metrics emergent from many-body interactions of various experimental systems. More recently, an exciting possibility of the co-existence of topological order with broken symmetry~\cite{Xia2011, Mulligan}, leading to the ``nematic" FQH effect, has also been proposed \cite{Sachdevnematic,chaikin, MaciejkoPhysRevB.88.125137, YouPhysRevX.4.041050, Jacksonnaturecomm}.  In this case, the nematic order arises due to spontaneous symmetry breaking, supported by recent numerical calculation~\cite{Regnault2016} and experiments using hydrostatic pressure~\cite{Samkharadze2016}.

For  a two-body interaction with rotational symmetry, i.e. $V(\bf{k}) = V(|k|)$, the understanding of different FQH states was greatly facilitated by the concept of pseudopotentials (PPs) introduced by Haldane~\cite{Haldane83, prangegirvin} in which the effective interaction is expanded by a complete basis $V_{\text{eff}}(k) = \sum_m c_m V_m(k)$.  The kernel function  is $V_m(k) = L_m(k^2) e^{-k^2/2}$ in which $L_m(k)$ are the Laguerrel polynomials. The index $m$ is the relative guiding center angular momentum of the two interacting particles. With these PPs, it is well known that some of the  FQH model wave functions are the zero energy ground state for the model Hamiltonians with few PPs, such as the $V_1$ hard-core Hamiltonian for the  $\nu = 1/3$ Laughlin state and  $V_1 + V_3$ Hamiltonian for $\nu = 1/5$ Laughlin state.  For a general two-body interaction,  we recently have developed~\cite{bo2016} the pseudopotential description generalized to cases without the assumption of  rotational symmetry. In this language, arbitrary two-body interactions can be decomposed in a matrix form.  The diagonal terms are the original Haldane's PPs for isotropic interaction, while the off-diagonal part describes the anisotropy.  We found some of the systems can be simply described by few PPs, such as the fractional Chern insulator with quadrupole interaction. Moreover, the anisotropic  PPs are found to be related to the nematic order in the nematic FQH phase. 

In this paper, as another application, we consider the spin polarized dipolar fermions in a fast rotating trap~\cite{mingwuPRL, Ni231}.  When the rotating frequency is approaching the trap frequency along $z$ axis, the Coriolis force experienced by the electrons leads to an effective perpendicular magnetic field. Therefore, the quantum Hall states are expected in the fast rotating limit~\cite{cooper08, RMP81647,PhysRevLett.94.070404,oster07, baranov08, Qiu2011} . However, the experimental realization of the FQH states in the rotation system has been elusive so far. The main problem is the precise control on the rotation frequency without crossing the rotational instability~\cite{PhysRevLett.92.040404}. Recently, other ways of engineering synthetic magnetic fields  to realizing the FQH states have been proposed, such as the strained optical lattice~\cite{PhysRevA.93.033640}, optical dressing~\cite{naturelin,PhysRevLett.93.033602} of atoms in continuum and laser-induced tunneling in optical lattices~\cite{PhysRevLett.111.185301, PhysRevLett.111.185302,Aidelsburger,Kennedy} . It is worth mentioning that a method of direct control over the total angular moment instead of  rotation frequency by using spin-flip induced insertion of angular momentum~\cite{PhysRevLett.110.145303} was proposed recently which circumvents the prime experimental difficulties toward the realization of the quantum Hall regime in harmonically trapped gases. 
\begin{figure}
 \includegraphics[width=5cm]{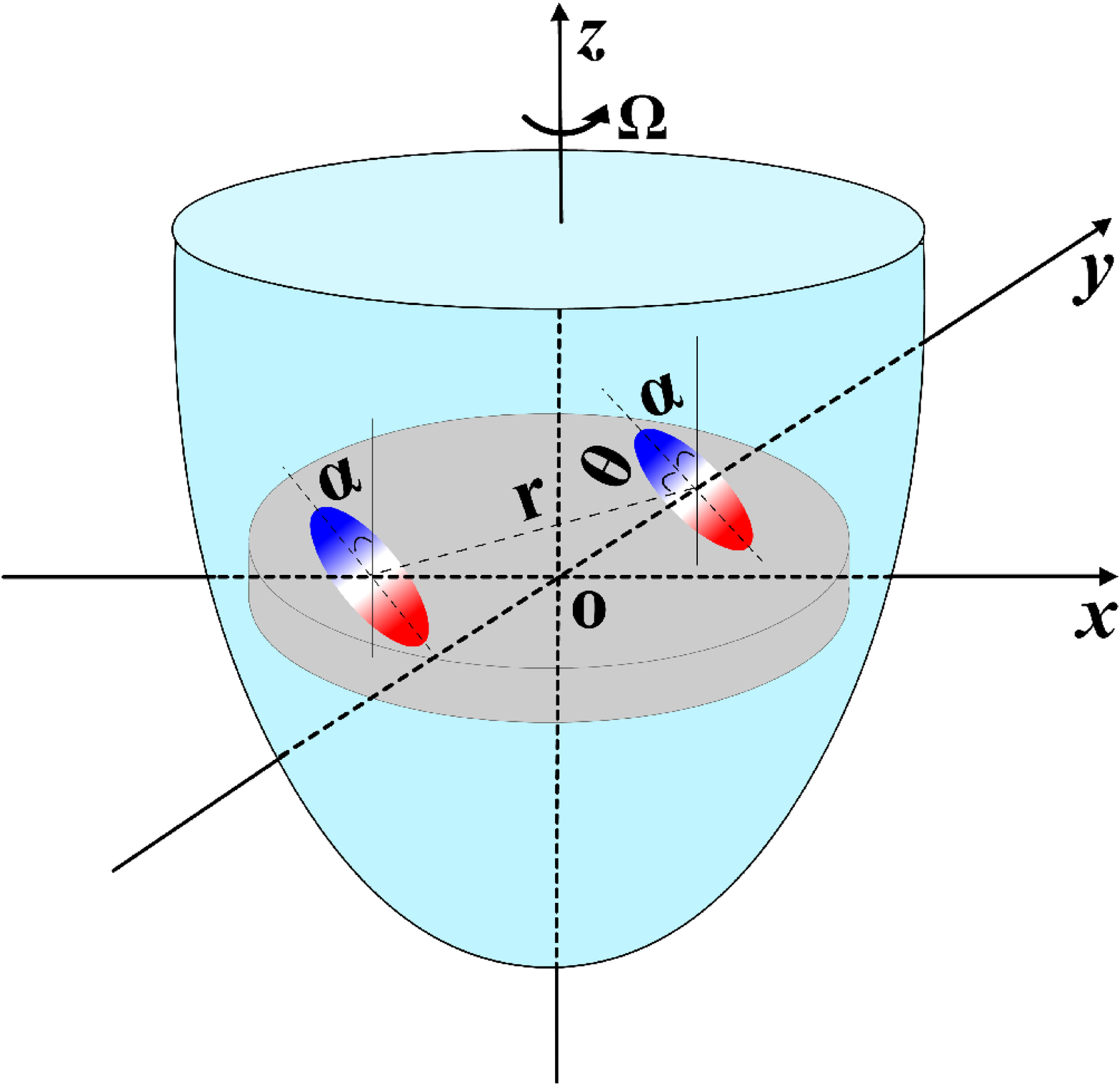}
 \caption{\label{model}The sketch map for the dipolar Fermions in a fast rotating harmonic trap. While the rotating frequency $\Omega$ is approaching to the trap frequency, the system is equivalent to the two-dimensional electron gas in a strong magnetic field. $\alpha$ is the angle between the direction of the dipole moment and $z$ axis.}
\end{figure}

In this work, we assume that the dipole momentum of all particles are polarized by an external orienting field which is at an angle $\alpha$ about the $z$ axis. By tilting the direction of the dipole moments with respect to the perpendicular axis, the dipole-dipole interaction can be tuned to be either isotropic or anisotropic. We characterize the dipole-dipole interaction systematically by the generalized pseudopotential especially when rotational symmetry is broken. By comparing with a family of the generalized Laughlin wave function parameterized by one metric, we can precisely locate the critical point of the quantum phase transition. For a specific oriented angle, an intrinsic metric is defined to describe the anisotropy of the system which can be obtained by maximizing the overlap between the ground state and the family of the Laughlin state~\cite{HaldaneGeometry} which is defined as $\Psi_L^{\nu=1/m}(g) = \prod_{i < j} [b_i^\dagger(g)-b_j^\dagger(g)]^m|0\rangle$~\cite{QiuPhysRevB.85.115308}. They are defined as the non-degenerate zero energy ground state of the model Hamiltonian where $b_j^\dagger(g)$ is obtained by a Bogoliubov transformation from the guiding center operator $b_j = (R_x - i R_y)/\sqrt{2}$ and $b_j^\dagger = (R_x + i R_y)/\sqrt{2}$ in the rotationally invariant case. The $g$ is a  unimodular metric of the ``area preserving'' deformation which can be defined as
 \begin{eqnarray}
 g=\left(
 \begin{array}{cc}
             \cosh2\theta + \sinh2\theta\cos2\phi & \sinh2\theta\sin2\phi \\
             \sinh2\theta\sin2\phi &  \cosh2\theta - \sinh2\theta\cos2\phi
            \end{array} \right).\nonumber
\end{eqnarray} 
The $\phi$ and $\theta$ are the rotation and stretching parameters respectively. By introducing the $g$ metric, a circular motion becomes to be elliptical.   On the other hand, before the FQH gap closes, we can treat this phase as a driven nematic FQH phase with breaking rotational symmetry by an external field. Since the nematic phase is diagnosed by the geometric response of the FQH state, the effect of tilting the dipolar angle  should also be reflected in the nematic order calculation.  

Our paper is organized as follows. Section II provides a compact derivation of the effective interaction for the dipolar fermions. The analysis of the interaction and its decomposition with the basis of the generalized PPs are also given. In section III, we discuss the properties of the $\nu = 1/3$ FQH state in the LLL. The intrinsic metric, wave function overlap and the nematic order are used to describe the phase transition. Sec. IV gives the results for the FQH state on 1LL, in which we find the $\nu = 2 + 1/5$ FQH state is more stable and has less anisotropy.  Conclusions and discussions are presented in Sec. V. 

\section{Effective interaction of the dipolar fermions in 2D}

In the Bose-Einstein condensation of $^{52}\text{Cr}$ atom~\cite{PhysRevLett.94.160401} or the degenerate quantum gas of $^{40}\text{K}^{87}\text{Rb}$~\cite{PhysRevLett.78.586}, the interaction can be described as dipole-dipole interaction with the $s$-wave  collisional interaction vanishing for spin polarized fermions.  We assume all the dipoles are polarized in the same direction and without loss of generality, say $x-z$ plane. The polarized dipole interaction (in unit of $d^2/(4\pi\epsilon_0 l^3)$, where $d$ is the dipole moment of the neutral atom and $\epsilon_0$ is the vacuum permittivity) is
\begin{equation} \label{vddr}
 V_{dd}(\bold{r},\alpha) = \frac{1-3(\bold{\hat{d}}\cdot\bold{\hat{r}})^2}{r^3} = \frac{r^2 - 3(z\cos\alpha + x\sin\alpha)^2}{r^5}
\end{equation}
with $\bold{\hat{d}}$ being the direction of the polarized dipole (parametrized by the angle $\alpha$) as sketched in Fig. ~\ref{model}.  $\alpha = 0^\circ$ is the rotational invariant situation in which all the dipoles are oriented along the $z$-direction and $\alpha \neq 0$ is the case that the dipole moment has component in the $x$-direction which breaks the rotational symmetry.

\begin{figure}
\includegraphics[width=8cm]{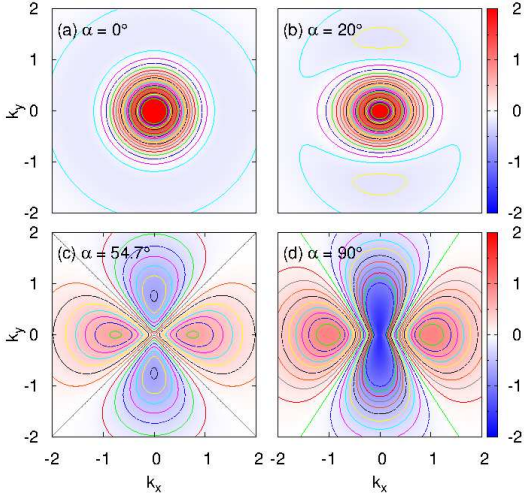}
\caption{\label{vk} The effective interaction function $V_{\text{eff}}(\bold{k})$ in the LLL for different tilted angles $\alpha$ at $q = 0.01$.}
\end{figure}
For a neutral particle  in a fast rotating limit in which the rotating frequency is close to that of the harmonic trap potential, if we assume the motion of the dipole in $z$-direction is frozen into its ground state, the quasi-two-dimensional single particle wave function can be written in a product form~\cite{cooper08}:
\begin{eqnarray}
 \psi_m(\vec{r}) = \frac{1}{\sqrt{\pi^{1/2}q}} e^{-z^2/2q^2} \frac{\rho^m e^{im\varphi} e^{-\rho^2/4}}{\sqrt{2\pi 2^m m!}}
\end{eqnarray}
where the three-dimensional vector is decomposed as $\vec{r} =(x, y, z) =  (\vec{\rho}, z)$. $q$ is defined as the thickness in $z$ direction in unit of $l = \sqrt{\hbar/(2\mu\omega)}$ where $\mu$ is the effective mass and $\omega$ is the frequency of the trap potential. Since the interaction only depends on relative part of the two-body wave function, for which the $z$-direction relative wave function is $\phi(z) = \frac{e^{-z^2/4q^2}}{(2\pi q^2)^{1/4}}$. The effective two-dimensional interaction can be obtained by integrating out the degree of freedom in $z$-direction:
\begin{eqnarray}\label{effectV}
V_{\text{2Deff}} (\bold{r}, \alpha)  = \int dz V_{dd}(\bold{r}, \alpha) |\phi(z)|^2  
\end{eqnarray}
where the general interaction $V_{dd}(\bold{r}, \alpha)$ can rewritten as
\begin{eqnarray}
 V_{dd}(\bold{r},\alpha)  = \frac{3\cos^2\alpha-1}{2}\frac{r^2 - 3z^2}{r^5} - \frac{3\sin^2\alpha \sin^2\theta\cos 2\varphi}{2r^3} \nonumber.
\end{eqnarray}
The first term is rotationally invariant, and when the dipoles are aligned along $z$ axis  $V_{dd}(\bold{r}, 0) = \frac{r^2 - 3z^2}{r^5}$.  The second term contributes  the anisotropic part of the interaction since it depends on the direction of the dipole $\{\theta, \varphi\}$. It is interesting to see that the isotropic interaction has a sign change at $3\cos^2\alpha-1 = 0$, i.e. the so-called magic angle $\alpha_c = 54.74^\circ$~\cite{cooper08}, at which the isotropic term vanishes and the system is completely determined by the anisotropic interaction.
According to Eq. (\ref{effectV}), we get the expression for the effective two-dimensional interaction:
\begin{eqnarray}
 V_{\text{2Deff}} (\bold{r}, \alpha) = \frac{e^{\rho^2/4q^2}}{2\sqrt{2\pi}q^5}[\mathcal{A}K_0(\frac{\rho^2}{4q^2}) + \mathcal{B}K_1(\frac{\rho^2}{4q^2})]
\end{eqnarray}
in which
\begin{eqnarray}
 \mathcal{A} &=&  (\rho^2 + 2q^2)\cos^2\alpha - (x^2 + q^2) \sin^2\alpha \nonumber \\
 \mathcal{B} &=& -\rho^2 \cos^2\alpha +  (q^2 + x^2 - 2q^2x^2/\rho^2) \sin^2\alpha \nonumber
\end{eqnarray}
with $K_l$ being the $l$'th modified Bessel function of the second kind. Then its Fourier transformation is given by
\begin{eqnarray}\label{v2ddefk}
 V_{\text{2Deff}} (\bold{k}, \alpha) &=& \frac{1}{2\pi}\int d^2\rho e^{i \bold{\rho}\cdot \bold{k}}   V_{\text{2Deff}} (\bold{r}, \alpha) \nonumber \\ 
 &=& \frac{4}{3q}\sqrt{\frac{\pi}{2}} [\mathcal{C} - \mathcal{D} \epsilon(\frac{qk}{\sqrt{2}})]
\end{eqnarray}
where
\begin{eqnarray} 
 \mathcal{C} &=& 3\cos^2\alpha - 1 \\
 \mathcal{D} &=& 3(\cos^2\alpha - \sin^2\alpha \cos^2\varphi_k)
\end{eqnarray}
and $\epsilon(x) = \sqrt{\pi}x e^{x^2} \text{erfc}(x)$.
The first term is a constant depends on $\alpha$, and the second term depends on the angle of the vector $\vec{k}$  ($\cos^2\varphi_k = \frac{k_x^2}{k_x^2 + k_y^2}$) which obviously breaks the rotational symmetry. After projecting to the Hilbert space of a single LL, two-body Hamiltonian is given by $\mathcal{H} = \frac{1}{2\pi} \int d^2 \bold{k} V_{\text{eff}}(\bold{k}) \rho(\bold{k}) \rho(-\bold{k}) $ with guiding center density operator $\rho(\bold{k}) = \sum_i e^{i k_a R_i^a}$ and the projected effective interaction
\begin{eqnarray}\label{veff}
 V_{\text{eff}}(\bold{k}, \alpha) = V_{\text{2Deff}} (\bold{k}) [L_N(|k|^2/2)]^2e^{-|k|^2/2}
\end{eqnarray}
for the electrons in the $N$'th Landau level.  In Fig.~\ref{vk}, we show the contour plot of the effective interaction $V_{\text{eff}}(\bold{k}, \alpha)$ for different dipole angles in the LLL.  Obviously, $V_{\text{2Deff}} (\bold{k}, 0)$ is rotational symmetric. For small $\alpha$ as shown in Fig.~\ref{vk}(b),  $V_{\text{2Deff}} (\bold{k}, \alpha)$ breaks into a $C_2$ symmetric shape and the contour lines become elliptic. At the magic angle, as shown in Fig.~\ref{vk}(c), the contour of the effective interaction has a quadrupolar  structure,
 similar to the $V_{m,2}$ PPs as shown in Fig.1 of the Ref.~\onlinecite{bo2016}.  In this case, the two-body interaction is repulsive in the $k_x$ direction and attractive in $k_y$ direction; thus we expect that all the dipoles align along in $x$ direction in real space energetically since the kinetic energy has been quenched by the effective magnetic field.  
 We thus expect the FQH state would undergo quantum phase transition into a compressible state with increasing $\alpha$.

For a two-body interaction without rotational symmetry, some of us recently found~\cite{bo2016} that a generalized pseudopotential description can be defined by:
\begin{eqnarray} \label{PPS}
 V^+_{m,n}(\bold{k}) &=& \lambda_n \mathcal{N}_{mn}(L_m^n(|k|^2) e^{-|k|^2/2} {\bold{k}}^n + c.c) \nonumber \\
 V^-_{m,n}(\bold{k}) &=& -i \mathcal{N}_{mn}(L_m^n(|k|^2) e^{-|k|^2/2} {\bold{k}}^n - c.c)
\end{eqnarray}
where the normalization factors are $\mathcal{N}_{mn} = \sqrt{2^{n-1}m!/(\pi(m+n)!)}$ and $\lambda_n = 1/\sqrt{2}$ for 
$n = 0$ or $\lambda_n = 1$ for $n \neq 0$. They satisfy the orthogonality
\begin{eqnarray} 
 \int V^\sigma_{m,n}(\vec{k}) V^{\sigma'}_{m',n'}(\vec{k}) d^2 k  = \delta_{m,m'}\delta_{n,n'}\delta_{\sigma,\sigma'}
\end{eqnarray}
thus the effective two-body interaction including the anisotropic ones can be expanded as
\begin{eqnarray} \label{ppformula}
 V_{\text{eff}}(\bold{k}) = \sum_{m,n,\sigma}^\infty c^{\sigma}_{m,n} V^{\sigma}_{m,n}(\bold{k})
\end{eqnarray}
with the coefficient
\begin{eqnarray}
 c^{\sigma}_{m,n} = \int d^2 k V_{\text{eff}}(\bold{k})  V^{\sigma}_{m,n}(\bold{k}).
\end{eqnarray}

The rotational invariant interaction only contributes the terms with $n=0$ and the $n \neq 0$ terms depict the anisotropy of the system. From Eq.~\ref{v2ddefk}, the anisotropic interaction comes from the term with $\cos^2\varphi_k = \frac{1}{2}[1 - \cos (2\varphi_k)]$ which only has non-zero contribution for the PPs with $n = 2$. The results for  any Landau level can be written in terms of the regularized hypergeometric function:
\begin{eqnarray} \label{cmn}
 c_{m,n}(\alpha) = \frac{1}{2}(3\cos^2\alpha - 1) c_{m,0} \delta_{n,0} + \sin^2\alpha c_{m, \pm 2} \delta_{n, \pm 2} \nonumber\\
\end{eqnarray}
 In LLL, the coefficients can be obtained analytically, where the diagonal terms are
\begin{eqnarray}
 c_{m > 0,0} &=& -\frac{\Gamma[m]}{2\sqrt{q}(q^2-2)}\{ _2F_1[\frac{1}{2},m, \frac{3}{2} + m, 1-\frac{2}{q^2}] \nonumber \\
 &-& {}_2F_1[\frac{3}{2},m, \frac{3}{2} + m, 1-\frac{2}{q^2}] \} \nonumber \\
 c_{0,0} &=& \frac{1}{6\sqrt{\pi}} \frac{\sqrt{\lambda}}{\lambda - 1}[ 2\lambda + 1 - 3 \lambda \frac{\tanh^{-1}\sqrt{1-\lambda}}{\sqrt{1-\lambda}}]
\end{eqnarray}
and the off-diagonal terms are:
\begin{eqnarray}
 c_{m, \pm 2} = \mathcal{F} {}_2F_1[\frac{1}{2}, m+1, \frac{7}{2}+m, 1-\frac{2}{q^2}]
\end{eqnarray}
with $\lambda = 2q^{-2}$ and $\mathcal{F} = -\frac{3\sqrt{(m+1)(m+2)}\Gamma[m+1]}{16\sqrt{2}q}$.

\section{Numerical results in the LLL}

\begin{figure}
\includegraphics[width=9cm]{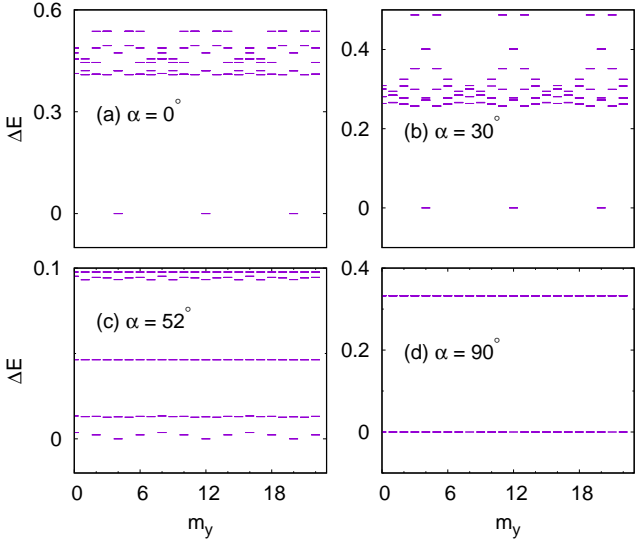}
\caption{\label{energyspectrum} Energy spectrum for 8 electrons in 24 orbitals on a torus at different tilted angles. The aspect ratio of the torus is set to be one.}
\end{figure}

Previously, most of the numerical investigation of the FQH state and its phase transition of this system were performed in disk geometry~\cite{oster07,baranov08,Qiu2011} which itself has rotational symmetry. It has drawbacks that the number of electron orbits should in principle be infinite (or finite with converging ground energy) while studying the anisotropic interaction. In addition,  the total angular momentum is not a good quantum number.  Thus the system size is limited (i.e., 6 electrons in ref.~\onlinecite{Qiu2011}) for the finite size diagonalization. Moreover,  in the disk geometry, the low-lying excited states are the gapless edge excitations which shift the ground state angular momentum as the trap potential is varied; therefore, the ground state overlap is not accurate for the determination of the phase transition in the bulk.  Further more, when tilting the dipoles, we previously~\cite{Qiu2011} calculate the ground state wave function overlap with respect to the rotational invariant Laughlin wave function $ |\Psi_0\rangle = \prod_{i<j}(z_i - z_j)^m e^{-\sum_k|z_k|^2/4}$. With the new geometric theory of the FQH states~\cite{HaldaneGeometry},  we know the anisotropic ground state may have large overlap with one of the family of the Laughlin states $|\Psi_{L}(g)\rangle$~\cite{QiuPhysRevB.85.115308} when it has small overlap with $|\Psi_{0}\rangle$.  Therefore,  it is useful to clarity that the overlap with the Laughlin state with the proper metric is a good diagnostic tool for detecting phase transition, while overlap with the isotropic Laughlin state is not as physically significant.

In this section, we perform the numerical diagonalization on torus geometry in which the translational symmetry of the system is conserved while all the dipoles are oriented in the same direction. The good quantum numbers are thus the translational momentum in the $x$ and $y$ direction, i.e., $k_x =m_x \frac{2\pi}{L_x} $, $k_y =m_y \frac{2\pi}{L_y} $. With these quantum numbers, we can go to larger system sizes. In addition, the compact torus geometry does not have edges and the phase transition can be determined based on if the ground state energy gap is closed. In Fig.~\ref{energyspectrum}, we plot the energy spectrum for 8 electrons in 24 orbitals at different tilting. When $\alpha = 0^\circ$, as we expected, the system has the feature of the typical $\nu=1/3$ Laughlin fractional quantum Hall state in which the ground states have a three-fold degeneracy due to the center of the mass translational symmetry~\cite{Haldane-PhysRevLett.55.2095} and are protected by a finite energy gap which makes the ground state incompressible. When increasing the tilting angle $\alpha$, the gap decreases monotonically as shown in Fig.~\ref{energyspectrum}(b). As $\alpha$ increases further and passes over the magic angle, some of the bulk states drop close to the ground states and finally the ground states have $N_{orb}$-fold degeneracy where $N_{orb}$ is the number of the orbitals. The energy gap is closed at $\alpha \simeq 53^\circ$ while $q = 0.01$. Here we should note that the energy gap closes at smaller $\alpha$ as increasing the layer thickness softens the interaction.

\begin{figure}
\includegraphics[width=9cm]{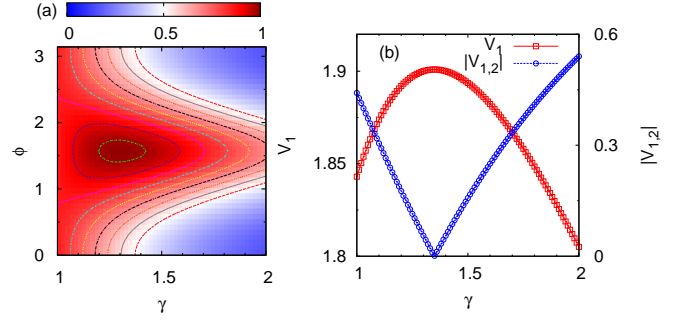}
\caption{\label{overlap}  (a)The ground state overlap $\mathcal{O}(\phi, \gamma)$ as varying the metric in the generalized Laughlin states. Here system is in a tilt with $\alpha = 40^\circ$ and $q = 0.01$. The maximum point at $\gamma = 1.3$ and $\phi = \pi/2$ corresponds to the intrinsic metric of the system. (b) The PPs $V_1^g$ and $|V_{1,2}^g|$ as a function of $\gamma$ at the same parameter as that in (a). The maximum $V_1^g$ corresponds to $V_{1,2}^g = 0$.  The $V_1^g$ is optimized at $\gamma \simeq 1.35$ which is close to the one for max($\mathcal{O}(\phi, \gamma)$) in (a).}
\end{figure}

 \begin{figure} 
\includegraphics[width=7cm]{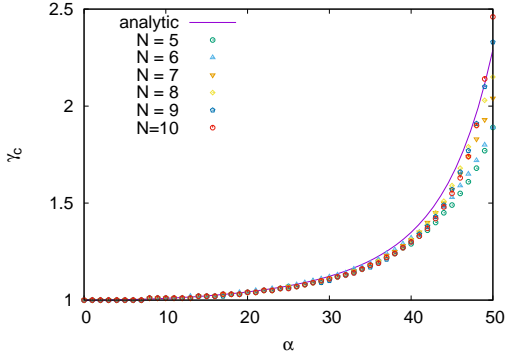}
\caption{\label{com} Comparison of the intrinsic metric for different systems and the analytic results for $q = 0.01$. The finite size results are very close to the analytic value for the thermodynamic limit. }
\end{figure}

We now show that overlap with the Laughlin states with the proper intrinsic metric can be used to detect phase transitions. This allows us to predict phase transitions from the ground state properties alone, without resorting to the energetics of the system. With the ground state wave function of the dipolar fermions for a given tilted angle $\alpha$, we compare two reference model wave functions. One is the rotational symmetric Laughlin $|\Psi_0\rangle$ which was used in previous study~\cite{Qiu2011}. Obviously, the overlap with this wave function $\mathcal{O}_0 = |\langle \Psi|\Psi_0\rangle|^2$ only tells us how far the anisotropic wave function is from the isotropic model wave function. 
With the knowledge of the generalized Laughlin states parameterized by a unimodular metric,  the exact phase boundary should be determined by calculating the overlap with running over all the generalized model wave functions $|\Psi_L(g)\rangle$ and picking out the maximum value, i.e., $\max(\mathcal{O}_g)$ where $\mathcal{O}_g= \mathcal{O}(\phi, \gamma) = |\langle \Psi|\Psi_L(g)\rangle|^2$.

 \begin{figure} 
\includegraphics[width=9cm]{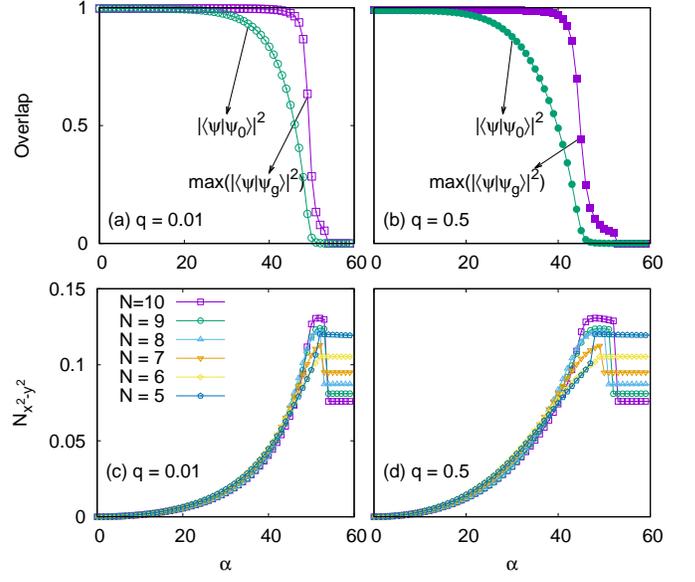}
\caption{\label{fig7} The overlap $\mathcal{O}_0$ and $\max(\mathcal{O}_g)$ as a function of the $\alpha$ for different layer thicknesses $q=0.01$(a) and $q=0.5$ (b). The system is for 10 electrons in 30 orbitals on the torus with aspect ratio one. (c) and (d) show the nematic order parameter $N_{x^2-y^2}$ as a function of the $\alpha$ for finite systems with 5-10 electrons at $\nu = 1/3$ filling in the LLL. }
\end{figure}

In our case, we restrict the direction of the dipoles in the $z-x$ plane, thus the parameter $\phi$ should be a constant. Therefore, we define a single parameter $\gamma = \cosh2\theta + \sinh2\theta$ to describe the metric as that in the study of the band mass anisotropy~\cite{BoYangPhysRevB.85.165318}. $\gamma = 1$ corresponds to the isotropic case with rotational symmetry. For a given ground state wave function of the  dipolar fermions $|\Psi\rangle$, to find out its intrinsic metric $\gamma_c$, we need to calculate all the overlap $\mathcal{O}_g$ and the intrinsic metric corresponds to the one with maximum overlap. Fig.~\ref{overlap} (a) shows the contour plot of the $\mathcal{O}(\phi, \gamma)$ for the system with $\alpha = 40^\circ$. It is shown that the maximum overlap corresponds to $\phi = \pi/2$ and $\gamma_c \simeq 1.3$. Thus for any tilted angle $\alpha$, to find out the intrinsic metric, we fix the $\phi = \pi/2$, and look for the maximum overlap with varying the stretching parameter $\gamma$.  In Fig.~\ref{com}, we plot the intrinsic metric $\gamma_c$ as a function of the tilted angle $\alpha$ for several finite size systems in the FQH regime. It is shown that the $\gamma_c$ is close to unity for small tilting and increases dramatically near the phase transition. 
In Fig.~\ref{fig7} (a) and (b), we plot two types of the overlap as a function of the $\alpha$ for two different thicknesses $q$. The results for two different layer thicknesses share the same conclusion that $\mathcal{O}_0$ is always smaller and drops faster than the $\max(\mathcal{O}_g)$. Thus if we define the phase transition boundary as the peak of the first order derivation, i.e., the peak of the fidelity,  using $\mathcal{O}_0$ will always over-estimate the critical point of the phase transition. This discrepancy increases with the layer thickness, and there is a region that $\max(\mathcal{O}_g)$ is almost one and $\mathcal{O}_0$ is very small. It means that the ground state is one of the $|\Psi_L(g)\rangle$  instead of the $|\Psi_0\rangle$. For the system with a larger thickness, both the $\mathcal{O}_0$ and $\max(\mathcal{O}_g)$ drop faster than the one with smaller thickness, which illustrates that the anisotropic effect is more prominent for thicker systems.
  
 Since all the dipoles are polarized in the same direction, and the anisotropic interaction has opposite sign in two perpendicular directions in the plane (see Fig. ~\ref{vk}), one would expect a partially melted solid with very large anisotropy. Inspired by the theory of classical liquid crystals, a nematic state is proximate to various phases with broken translational symmetry (i.e.stripe or smectic phases). In our case, the mechanism of the symmetry broken is driven by an external field. However, we can also borrow the concept of the nematic phase to discuss symmetry-breaking and phase transition in the dipolar system. In the torus geometry,  the Ising nematic order parameter~\cite{Regnault2016} is defined as
\begin{eqnarray}
 N_{x^2 - y^2} = \sum_{\bold{k}} (\cos k_x - \cos k_y) \langle \Psi | \rho(\bold{k}) \rho(-\bold{k})|\Psi\rangle,
\end{eqnarray}
which is actually a microscopic observable with quadrupolar symmetry. The numerical results are depicted in Fig.~\ref{fig7} (c) and (d) for two different layer thicknesses.  For small tilting, the nematic order parameter $N_{x^2 - y^2}$ for different system sizes almost collapse on the same curve. It smoothly increases as a function of the tilting angle $\alpha$ in the FQH phase, or before the gap closing.  It is interesting to see that the $N_{x^2 - y^2}$ has an inflection point when the isotropic overlap  $\mathcal{O}_0$ drops to zero and another inflection when $\max(\mathcal{O}_g)$ goes to zero. The window between the two inflection points becomes wider as increasing the layer thickness. In this window, the nematic order seem to be saturated and does not increase as rapidly as that when $\mathcal{O}_0$ is nonzero.  
Therefore, we conclude that the nematic order parameter is a character of breaking the rotational symmetry of the system.The sudden drop and being a constant for large $\alpha$ can be understood as follows. In the dipolar compressible phase, all the particles attract with each other and form a dipolar cluster, i.e., all the particles occupy $N_{orb}$ adjacent orbitals (one orbital can host only one particle because of the Pauli exclusive principle for fermions).  Therefore, the discontinuous drop of the nematic order parameter is related to a sudden change of the area of the many-body system which can be diagnosed as the critical point.

\begin{figure}
\includegraphics[width=8cm]{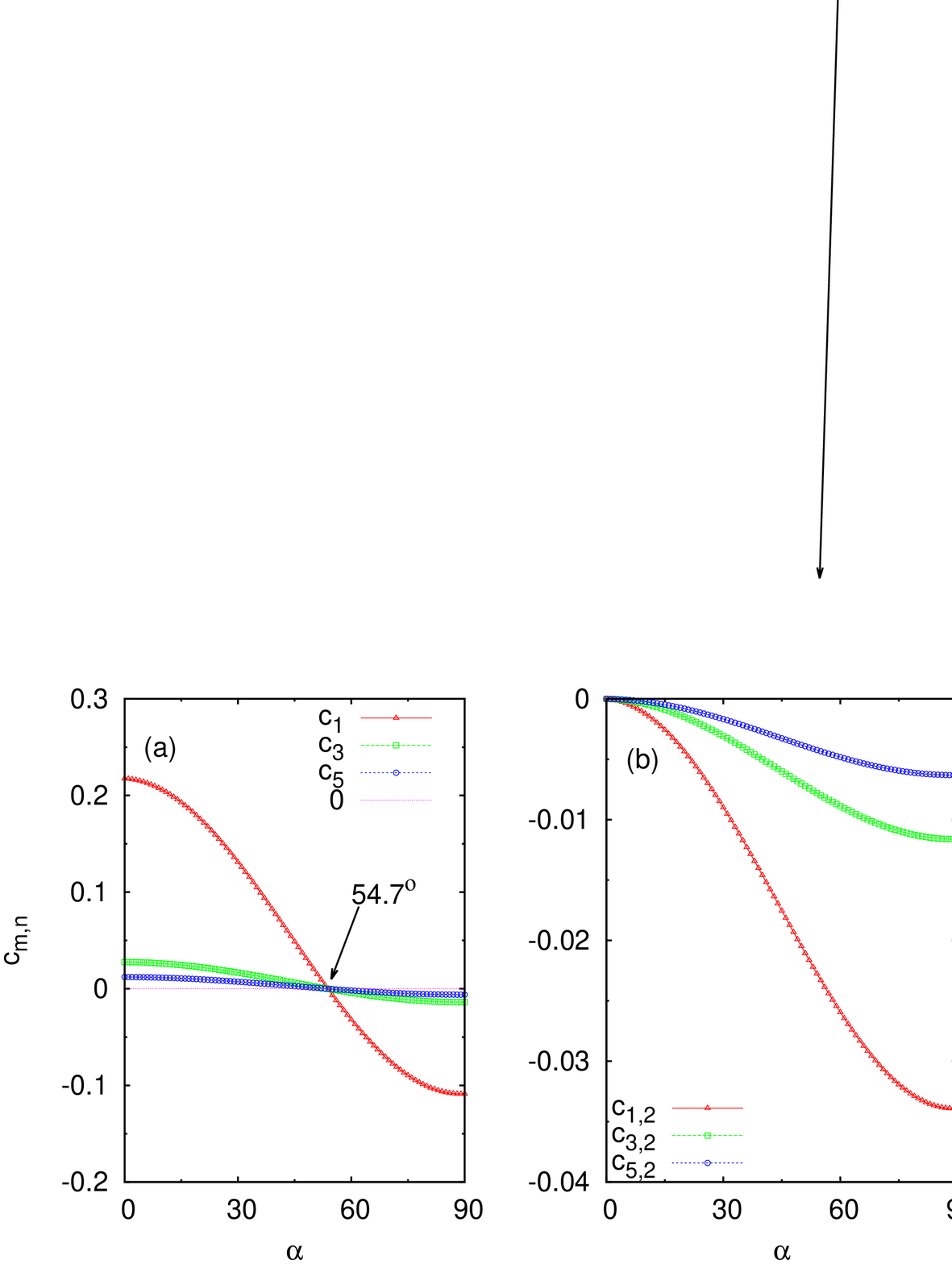}
\caption{\label{pps} PPs as a function of the tilted angle $\alpha$ for the LLL.   (a) and (b) show the three dominant diagonal and off-diagonal terms with odd $m$. At the magic angle $\alpha=54.74^\circ$,  all the diagonal terms vanish.}
\end{figure}

In order to investigate this phase transition from the fundamental two-body interaction, as was analyzed in Sec. II,  we plot a few dominated PPs for the LLL as being formulated in Eq. (\ref{cmn}) with odd $m$ (for fermions) by varying the tilted angle $\alpha$ for $q = 0.01$ in Fig.~\ref{pps}. One feature is that the coefficient of the $V_1$ potential, $c_1$, is one order larger than the others while $\alpha = 0^\circ$ which means the Laughlin state at $\nu = 1/3$ can be very stable. The magnitude of the diagonal coefficients monotonically decrease and have a sign change at the magic angle when varying the tilted angle $\alpha$. Therefore, a phase transition is expected when the  dipole direction is rotated from $z$ to $x$. Moreover, the diagonal terms are zero at the magic angle and the Hamiltonian only contains the anisotropic interaction. In particular, the coefficient of the anisotropic PP $V_{1,2}$, $c_{1,2}$,  is approximately three  times larger than the second largest one $c_{3,2}$. Therefore, we expect that the anisotropy of the system is mainly determined by the $V_{1,2}$ pseudopotential in the LLL.  A simple model Hamiltonian for this system can be written as $H = V_1  + \lambda_0 V_{1,2}$ where the parameter $\lambda$ depends on the tilted angle $\alpha$ and the layer thickness $q$. When $\alpha = 90^\circ$, as shown in Fig.~\ref{pps}, the interaction is dominated by negative $V_1$ and $V_{1,2}$. In this case, the dipole-dipole interaction is fully attractive; thus all the dipoles likely bound together and occupy a series of adjacent orbits is expect in the compressible ground state.  

Since the Laughlin state $|\Psi_{L}(g)\rangle$ is the exact zero energy ground state of the $V_1^g$ Hamiltonian, all the other PPs in the realistic interaction can be treated as perturbations. Therefore, we suspect that maximizing the overlap, or the intrinsic metric corresponds to maximizing the coefficient of the $V_1^g$~\cite{botitleB}. In Fig.~\ref{overlap}(b), we plot the $V_1^g$ and $|V_{1,2}^g|$  as varying $\gamma$ with the same parameter as that in Fig.~\ref{overlap} (a). It is shown that the $\gamma$ with maximum $V_1^g$ is exactly the one corresponding to $V_{1,2}^g= 0$.  Comparing with Fig.~\ref{overlap} (a), we find that the metric $\gamma_c$ for optimizing the overlap and $V_1^g$ are almost the same ($\gamma_c \simeq 1.35$ for $\alpha = 40^\circ$).
From Eq.(\ref{PPS}), we find the following relation should be satisfied:
 \begin{eqnarray}
 \partial_\theta V_1^g = -\frac{\sqrt{3}}{2}(\cos\phi V_{1,2}^{g}-\sin\phi V_{1,2}^{g}).
\end{eqnarray}
Therefore, from this condition, we can obtain the analytic result for the intrinsic metric in the thermodynamic limit which is also plotted in Fig.~\ref{com} as a comparison. It is shown that the numerical results for larger system size are getting close to the analytic one.
 
\section{dipolar fermions in the 1LL}

 \begin{figure}
\includegraphics[width=8cm]{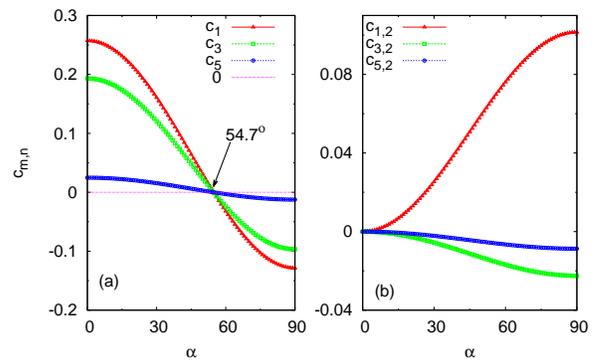}
\caption{\label{pps1ll} The same plot as in Fig.~\ref{pps} for the 1LL.}
\end{figure}

\begin{figure}
\includegraphics[width=9cm]{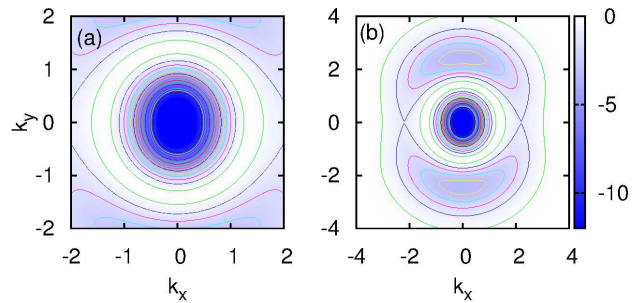}
\caption{\label{vkA90} The $V_{\text{eff}}(\bf{k})$ for 1LL while $\alpha = 90^\circ$.  In (a), we can see the squeeze direction of the contours  rotated $90^\circ$, however, in a large $|k|$ plot as shown in (b), the stretching direction recovers back as that in the LLL.}
\end{figure}

In this section, we focus on the FQH state of the dipolar fermions in the 1LL which only has a form factor difference from that of the LLL in the effective interaction.   After substituting the form factor in Eq.~\ref{ppformula} and performing the numerical integration, we obtain the generalized PPs. Some of the dominant ones are plotted in Fig.~\ref{pps1ll} as a function of the $\alpha$. In contrast to the LLL, we find the PPs for $V_1$ and $V_3$ are comparable and one order larger than the others in the 1LL. For example, $\{c_1, c_3, c_5 \cdots\} = \{0.2572, 0.1932, 0.0248 \cdots\}$ with $q = 0.01$ and $\alpha = 0^\circ$. Therefore, we expect that the $\nu = 2 + 1/5$ Laughlin state, which is the exact zero energy state for model hamiltonian $H = V_1 + V_3$, should be the most stable ground state in the 1LL for dipolar fermions. It is somewhat counterintuitive since we generally expect the $\nu = 2 + 1/3$ Laughlin state is stable for Coulomb interaction.  Therefore, we know the $\frac{1}{r^3}$ short range dipole-dipole interaction results in a comparable $V_3$ components in the PPs in 1LL. Similar to the case in the LLL, all the isotropic PPs decrease with increasing tilted angle and have a sign change at the magic angle $\alpha_c$. The $c_1$ and $c_3$ have comparable values and one order larger than $c_5$ for any tilted angle. 

It is more interesting to look at the anisotropic PPs. Similar to the case in  LLL, the dominate anisotropic PPs is the $V_{1,2}$. The difference is that the coefficient of the $V_{1,2}$ in the 1LL has an opposite sign to that of the LLL.
All the other $V_{m>1,2}$ are negative both in the LLL and 1LL.  For a model Hamiltonian which only contains the $V_{1,2}$ anisotropic PPs, we know the sign change of the $V_{1,2}$ means a $90^\circ$ rotation of the metric.  In Fig.~\ref{vkA90},  we plot the $V_{\text{eff}}(\bf{k})$ in the 1LL with $\alpha = 90^\circ$. It is shown that the squeeze in the contours of the $V_{\text{eff}}(\bf{k})$ only has a $90^\circ$ rotation in a medium $|\bf{k}|$ region. From the overall plot of the $V_{\text{eff}}(\bf{k})$ as shown in Fig.~\ref{vkA90}(b), the dipoles still has the same stretching as that in the LLL. It has attractive interaction along $k_y$, or $x$ direction in real space. The intrinsic metric calculation is similar to that of the LLL as shown in Fig.\ref{overlap} which also confirms the discussions above. Therefore,  despite the smallness $V_{3,2}$, which has negative coefficient, it is not negligible.  The model Hamiltonian in the 1LL should be described by  $ H = V_1 + \lambda_1 V_3 + \lambda_2 V_{1,2} + \lambda_3 V_{3,2}$ with all the $\lambda_i$ depending on the $\alpha$ and $q$.

\begin{figure}
 \includegraphics[width=9cm]{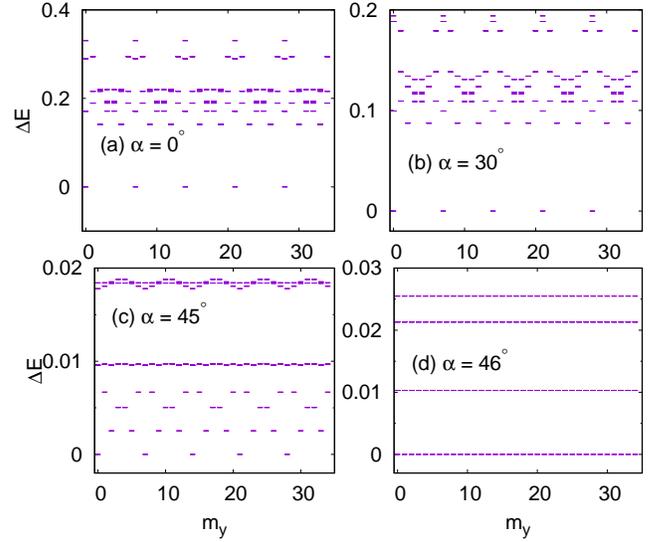}
 \caption{\label{energyspec1LL} The energy spectrum for 7 electrons at $\nu = 1/5$ on the 1LL for different dipole angle $\alpha$ when $q = 0.01$.}
\end{figure}
\begin{figure} 
\includegraphics[width=9cm]{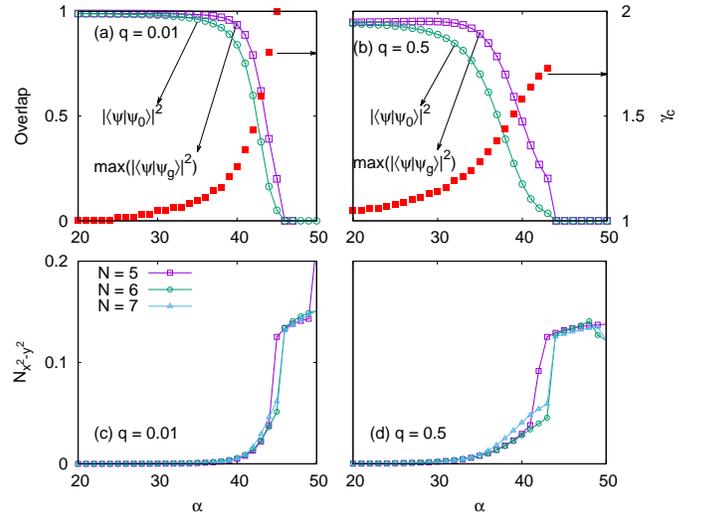}
\caption{\label{fig11}   Similar to Fig.\ref{fig7}. The comparison of the overlap and intrinsic metric are for the system with 7 electrons at $\nu = 2 + 1/5$ in the 1LL.}
\end{figure}

Fig.~\ref{energyspec1LL} depicts the energy spectrum for 7 electrons at filling $\nu = 2 + 1/5$ for different $\alpha$ when $q = 0.01$.  As being expected, the isotropic FQH state at $\alpha = 0^\circ$ has 5-fold degeneracy. The ground state energy gap survives until $\alpha = 45^\circ$. While $\alpha > 45^\circ$, as shown in Fig.~\ref{energyspec1LL}(d), the 5-fold degenerate ground states become to be $N_{orb}$-fold and therefore the system enters the compressible phase. The critical angle is smaller than that of the LLL and far from the magic angle which illustrates that the FQH in the 1LL is more fragile against the introduction of anistropy. In Fig.~\ref{fig11}(a) and (b) we perform the similar comparison of the wave function overlap $\mathcal{O}_0 $ and $\max(\mathcal{O}_g)$, the intrinsic metric $\gamma_c$ and different layer thicknesses.  Here the trial wave function $\Psi_g$ is obtained by diagonalizing the model Hamiltonian $H = V_1^g + V_3^g$.  (c) and (d) show the results of the nematic order parameter as a function of the $\alpha$.   Combining all the plots and comparing with the results in LLL as shown in Fig.~\ref{fig7},  we find that the difference between $\mathcal{O}_0$ and $\max(\mathcal{O}_g)$ is smaller in the 1LL for the same layer thickness. Taking $q = 0.01$ as an example,  in the LLL, there is a window $\alpha \in [30^\circ, 50^\circ]$ in which $\mathcal{O}_0$ drops very fast while $\max(\mathcal{O}_g)$ keeps close to one; and in the 1LL,  they have the same behavior as varying $\alpha$ although  $\max(\mathcal{O}_g)$ is always larger than $\mathcal{O}_0$.  This phenomenon tells us that the optimized $|\Psi_g\rangle$ is very close to $|\Psi_0\rangle$ in the 1LL before the phase transition, or tilting the dipolar angle $\alpha$ introduces less rotational symmetry broken in the 1LL comparing to that in the LLL. This can also be verified in the nematic order calculation as shown in Fig.~\ref{fig11} (c).  In the 1LL, the nematic order  is very close to zero until $\alpha \sim 40^\circ$. However, in the LLL as shown in Fig.~\ref{fig7} (c), the nematic order starts to become appreciatable even for small tilting angle.  The analysis for the case of $q = 0.5$ is the same. This is understandable, since in the LLL, the intrinsic metric of the $1/3$ Laughlin state is strongly modified by $V_{1,2}$, while in the 1LL, $V_{1,2}$ does not modify the intrinsic metric of the $1/5$ ground state. The latter is only affected by the magnitude of the coefficient of $V_{3,2}$, which is much smaller as compared to that of $V_{1,2}$.

\section{Conclusions and discussions}

\begin{figure}
\includegraphics[width=8cm]{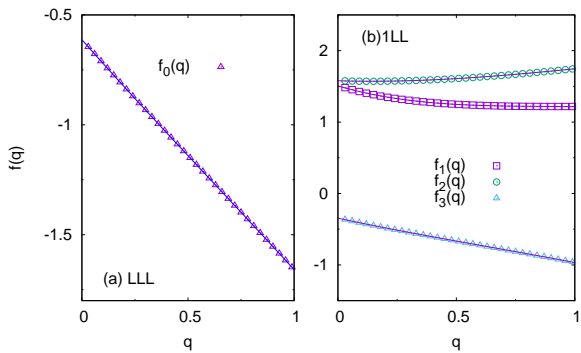}
\caption{\label{modelp}The fitted parameters for the model Hamiltonian for the (a) LLL  and (b) 1LL respectively.  The model Hamiltonian is written as $H = \text{sgn}(\alpha-\alpha_c)(V_1^g + \lambda_0 V_{1,2}^g)$ for the LLL and $H = \text{sgn}(\alpha-\alpha_c)(V_1^g + \lambda_1 V_3^g + \lambda_2 V_{1,2}^g  +\lambda_2 V_{3,2}^g)$ for the 1LL. The parameter $\lambda_i(\alpha, q)$ can be written as $\lambda_i(\alpha, q) = h(\alpha)f_i(q)$ where  
 $h(\alpha) = \frac{\sin^2(\alpha)}{\frac{1}{2}(3\cos^2(\alpha)-1)}$ and $f_i(q)$ are fitted as  $f_0(q) \simeq -0.61-1.1q-0.10q^2 + 0.05q^3$, $f_1(q) \simeq
1.5 - 0.95q + 1.09q^2-0.42q^3$, $f_2(q) \simeq 1.58 - 0.1q + 0.37q^2-0.1q^3$
 and $f_3(q)  \simeq -0.35-0.71q+0.2q^2-0.11q^3$ in the range $q \in [0,1]$.}
\end{figure}

In conclusion, we systematically study the FQH state and its phase transition in the fast rotating dipolar fermionic system. When the dipole moment is polarized and rotated by an external field, the interaction between two dipoles breaks the rotational symmetry.  We expand the anisotropic interaction in the basis of the generalized PPs.  As increasing the tilted angle $\alpha$, the isotropic PPs decrease monotonously  and have a sign change from repulsive to attractive at the magic angle $\alpha_c = 54.7^\circ$ and the amplitude of the anisotropic PPs are enlarged.  All the anisotropic PPs are negative except the $V_{1,2}$ in the 1LL.  Based on the analysis of the dominant PPs, we find that the most stable FQH states in the LLL and 1LL are $\nu = 1/3$ and $\nu = 2 + 1/5$ Laughlin FQH states respectively in the isotropic case.  When $\alpha > 0^\circ$, these FQH states are still robust, but the energy gap of the ground state becomes weaker.  After the gap closing, the ground state becomes $N_{orb}$-fold degenerate compressible states.

Since the interaction was decomposed by the generalized PPs and the anisotropy of this system is characterized by few anisotropic PPs,  one can construct simple model Hamiltoians to effectively characterize the incompressible states for the dipolar fermions. In the LLL, for the $1/3$ Laughlin state, the model Hamiltonian is given by $H = \text{sgn}\left(\alpha-\alpha_c\right)\left(V_1 +  \lambda_0 V_{1,2}\right)$. Similarly, in 1LL we have $H =\text{sgn}\left(\alpha-\alpha_c\right)\left( V_1 + \lambda_1 V_3 + \lambda_2 V_{1,2}  + \lambda_3 V_{3,2}\right)$ where in both cases we normalize the coefficient of the $V_1$ PPs to be one.  All the parameter $\lambda$'s depend both on the tilted angle $\alpha$ and the layer thickness $q$. From Eq. (\ref{cmn}), the $\alpha$ dependence is universal. If we define $\lambda_i(\alpha, q) = h(\alpha)f_i(q)$, we have $h(\alpha) = \frac{2\sin^2(\alpha)}{3\cos^2(\alpha)-1}$. The $q$ dependent function $f_i(q)$ is non-universal, though in the LLL it can be computed analytically. For thickness that is smaller as compared to the effective magnetic length, $f_i\left(q\right)$ can be Taylor expanded, and the results are shown in Fig.~\ref{modelp}.  These model Hamiltonians provide us a simple model to analyze the effects of the anisotropy (in particular the intrinsic metric) for the dipolar fermions in the FQH regime.

In order to describe the anisotropy of the dipole-dipole interaction, we calculate the intrinsic geometric metric $\gamma_c$ of the system. For the $\nu = 1/3$ Laughlin-like state in the LLL, $\gamma_c$ in the thermodynamic limit can be obtained analytically via maximizing the $V_1^g$ pseudopotential. We find the numerical results are consistent with the analytic ones.  For the $\nu = 2 + 1/5$ state in the 1LL, since there are two dominant PPs $V_1$ and $V_3$, it is still an open question on how we can attain the intrinsic metric analytically. In the geometric description of the FQH state, there is a family of the Laughlin wave functions which are zero energy state of the $H = V_1^g$. It is possible that the ground state for $\alpha > 0^\circ$ is well described by one of the $|\Psi_g\rangle$, although it may have very small overlap with the isotropic model wave function $|\Psi_0\rangle$. The boundary of the phase transition from FQH to the compressible state can be determined by the optimized wave function overlap $\max(\mathcal{O}_g)$ instead of  the $\mathcal{O}_0$ as being studied previously on disk. We find that the behaviors of the $\max(\mathcal{O}_g)$ are indeed consistent to the other order parameters of the phase transition, such as the energy gap,  intrinsic metric and the nematic order parameter.   On the other hand, an intriguing coexistence of the topological state with broken rotational symmetry leads to the nematic FQH order. Here, although the symmetry breaking is triggered by external field, in the resulting nematic FQH phase before the gap closing, we expect that the nematic order parameter $N_{x^2-y^2}$ can still be useful in describing the anisotropy in the FQH regime.  Our numerical calculations reveal that the $N_{x^2-y^2}$  at $\nu=1/3$ emerges soon after the direction of the dipole moment deviating the $z$ axis. However, for the $\nu = 2 + 1/5$ state, the $N_{x^2-y^2}$ stays at zero for a considerable tilting.  It tells us that the effect of the anisotropy in the 1LL is smaller than that in the LLL in this system. This can be understood as the $V_{1,2}$ has opposite sign to the other anisotropic PPs, which means the metric for $V_{1,2}$ has a $90^\circ$ rotation comparing to others. The mutuallyl-orthogonal metrics partly cancel the anisotropy of the system. Therefore, with the same tilted angle, the FQH in the 1LL has less anisotropy than that in the LLL.

\acknowledgements
This work is supported by National Natural Science Foundation of China Grants No. 11674041, No. 91630205,  Fundamental Research Funds for the Central Universities Grant No. CQDXWL-2014-Z006 and Chongqing Research Program of Basic Research and Frontier Technology Grant No. cstc2017jcyjAX0084. R-Z. Qiu is supported by NSFC No. 11404299. B. Yang is supported in part by Singapore A*STAR SERC "Complex System"  Research Program grant 1224504056.

\end{document}